\documentclass[prd,twocolumn]{revtex4}
\usepackage{graphicx, epsfig}
\usepackage{color}
\usepackage{mathrsfs}
\usepackage{amsmath}
\usepackage{amsfonts}
\usepackage{bm}

\newcommand{\be}{\begin{equation}}
\newcommand{\ee}{\end{equation}}
\newcommand{\bea}{\begin{eqnarray}}
\newcommand{\eea}{\end{eqnarray}}
\newcommand{\ba}{\begin{eqnarray}}
\newcommand{\ea}{\end{eqnarray}}

\newcommand{\gapp}{\mathrel{\raise.3ex\hbox{$>$}\mkern-14mu
              \lower0.6ex\hbox{$\sim$}}}
\newcommand{\lapp}{\mathrel{\raise.3ex\hbox{$<$}\mkern-14mu
              \lower0.6ex\hbox{$\sim$}}}

\begin{document}
\title{Cosmic Rays from Cosmic Strings with Condensates}
\author{Tanmay Vachaspati}
\affiliation{
Institute for Advanced Study, Princeton, NJ 08540\\ 
and\\
CERCA, Department of Physics, 
Case Western Reserve University, Cleveland, OH 44106-7079.\\
}

\begin{abstract}
\noindent
We re-visit the production of cosmic rays by cusps on cosmic 
strings. If a scalar field (``Higgs'') has a linear interaction
with the string world-sheet, such as would occur if there
is a bosonic condensate on the string, cusps on string loops 
emit narrow beams of very high energy Higgses which then decay 
to give a flux of ultra high energy cosmic rays. The ultra-high 
energy flux and the gamma to proton ratio agree with observations 
if the string scale is $\sim 10^{13}$ GeV. The diffuse gamma ray 
and proton fluxes are well below current bounds. Strings that are 
{\it lighter} and have linear interactions with scalars 
produce an excess of direct and diffuse cosmic rays and are ruled 
out by observations, while heavier strings ($\sim 10^{15}$ GeV) are 
constrained by their gravitational signatures. This leaves a narrow 
window of parameter space for the existence of cosmic strings
with bosonic condensates.
\end{abstract}

\maketitle

The detection of cosmic topological defects can provide a direct 
window to fundamental physics and the very early universe (for a 
review, see \cite{VilenkinShellard}). Hence it is no surprise 
that there has been a concerted effort to examine observational 
consequences of cosmic topological defects, including gravitational 
wave emission, gravitational lensing, fluctuations in the cosmic
microwave background, and ultra-high energy cosmic rays. If cosmic 
strings are superconducting, they may also lead to electromagnetic 
phenomena such as gamma ray and radio bursts. 

\begin{figure}
\scalebox{0.80}{\includegraphics{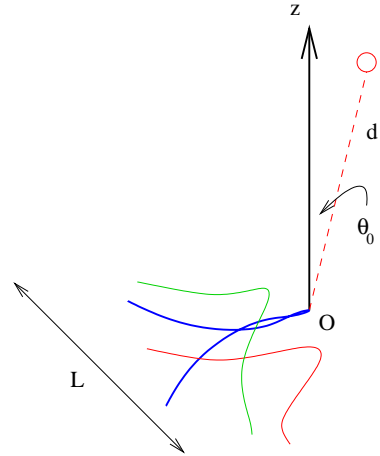}}
\caption{String segment at 3 different times,
with a cusp at O, with velocity along $z$. The size of the
curved section of string is $L$. The observer is at an
angle $\theta_0$ from the z-axis at a distance $d$. There
is strong emission from the cusp in a beam along the
$z$ axis.}
\label{schematics}
\end{figure}

The gravitational effects of cosmic strings are stronger for heavier 
strings and current observations of the cosmic microwave background 
rule out strings above the $\sim ~ {\rm few} \times 10^{15}$ GeV 
energy scale. As lighter strings are considered, gravitational 
effects become less significant, and other particle physics 
signatures become relatively important. If the strings are
superconducting, the currents on the string can lead
to electromagnetic radiation that could be observable.
However, if the strings are not superconducting,
such signatures will be absent, and one must turn 
to particle emission from cosmic strings. If string 
dynamics forces loop production at the smallest 
possible scales, as suggested in \cite{Vincent:1997cx},
particles can be copiously emitted, leading to strong
constraints. However, other studies indicate
that string loops are large compared to microscopic 
length scales \cite{VilenkinShellard} and particle emission 
is suppressed \cite{Vachaspati:1984yi}. Even in this case, 
portions of a string loop may get boosted to very high 
Lorentz factors, creating a ``cusp'' on the string 
(see Fig.~\ref{schematics}), and this may potentially 
provide a burst of particles that could be seen in cosmic 
ray detectors. 

Particle emission from cusps on cosmic strings has
been studied by several authors. Srednicki and Theisen
\cite{Srednicki:1986xg} considered a quadratic interaction 
of a scalar field with an idealized (zero thickness) string 
and came to the conclusion that particle emission is 
insignificant for astrophysical size string loops.
Our analysis for the radiation is similar to that of 
Refs.~\cite{Srednicki:1986xg,Damour:1996pv}, though the
particular {\it linear} interaction of the scalar field 
(call it $H$) with the string world-sheet, as would occur
if $H$ condenses on the string
\footnote{The condensate is similar to that for bosonic 
superconducting strings \cite{Witten:1984eb} but dissimilar 
in that we do not require charged modes to propagate along 
the string.}, has not been considered before. A linear 
interaction causes an enhancement of particle production 
by a factor of $M/m$ where $M$ is the string scale and $m$ 
is the mass of $H$. If $M$ is the Grand Unification scale, 
while $m$ is the electroweak scale, this factor can be as 
large as $10^{13}$.

Particle emission from cusps on {\em thick} strings has 
been considered in Ref.~\cite{earlyworks,Bhattacharjee:1989vu}. 
Now the cusp consists of overlapping, oppositely oriented 
strings that can annihilate and give off energetic 
particles. A careful study of this process, including 
numerical evolution of the field equations, shows that 
the resulting flux of particles is too small to be of
interest \cite{Gill:1994ic,BlancoPillado:1998bv}.
In contrast, the linear interaction mechanism we study 
is insensitive to the thickness of the string, and
occurs over a string length that is much larger
than the length over which cusp annihilation occurs.
Thus we can ignore cusp annihilation and work in
the zero thickness limit. 

There are important observational constraints on
the so-called ``top down'' models for production 
of ultra-high energy cosmic rays (UHECR), in which
a heavy particle decays to give ultra-high energy 
protons and gamma rays. The constraint arises because 
there are bounds on the diffuse gamma ray flux in 
the EGRET window and also on the fraction of photons 
to protons in the observed UHECR.
In previous studies, using cusp annihilation 
on cosmic strings as the source of Higgs injection, it was 
found that the EGRET constraint on diffuse gamma ray fluxes
implies that the flux of UHECR is uninterestingly low. 
However, the constraint on diffuse gamma rays is 
sensitive to the spectral features of the injected 
Higgs particles and to the particular interaction of the 
string. The Higgs particle emission that we consider yields 
a diffuse flux of gamma rays and protons that is below
EGRET bounds if the string scale is $> 10^6$ GeV and
an UHECR flux consistent with observations for
string scale $\sim 10^{13}$ GeV. The ratio of photons
to protons in the UHECR flux is also consistent with
current bounds. Hence our model can explain the observed 
UHECR and not run into trouble with the EGRET bounds.

It is to be noted that cosmic ray production by strings
increases as the string scale {\it decreases}. Hence
strings on scales less than $\sim 10^{13}$ GeV, and with
these interactions, are {\it excluded} by the observed flux
of UHECR. This is especially interesting since heavier
strings are constrained by their gravitational signatures.
Hence there is a narrow window between, say, 
$10^{13}-10^{15}$ GeV for the mass scale of cosmic strings 
having linear interactions with a scalar field.

To summarize, the novelty in the present work is that
we are considering a new, generic, interaction of cosmic
strings with scalar fields that leads to high energy 
particle emission. This interaction seems to have been 
missed in the literature. Also, we have been careful to
take the beamed nature of the emission into account.
We have focused on deriving approximate analytical 
estimates so as to keep the physical aspects of the
problem as apparent as possible. More detailed predictions 
will require numerical evaluation of the production
and propagation.

We begin in Sec.~\ref{fieldtheory} by describing the 
field theory interactions under consideration.
We then evaluate the rate at which a single cusp
emits Higgs particles in Sec.~\ref{higgsemission}. 
In Sec.~\ref{higgsinjection} we use the 
results of Sec.~\ref{higgsemission} to determine
the cosmological Higgs injection function. The diffuse 
gamma ray and proton fluxes are calculated in 
Secs.~\ref{cascade} and \ref{protonflux} respectively. 
These cosmic rays originate in strings that are relatively 
far away from the Earth. Higgs emission from strings 
that are closer to us and pointed at us can give UHECR. 
The direct flux of UHECR is calculated in 
Sec.~\ref{directflux} where we also discuss the ratio 
of photons to protons in UHECR.  We conclude in 
Sec.~\ref{conclusions}. In Appendix~\ref{loopdynamics} 
we summarize some known facts about cosmic string loop 
dynamics.
 
\section{Field theory}
\label{fieldtheory}

The interaction we consider is
\begin{equation}
S_{\rm int} = - \kappa M \int d^2\sigma \sqrt{-\gamma} ~ h 
\end{equation}
where $\kappa$ is a coupling constant assumed to be
$\sim 1$, $M$ is the string
energy scale assumed to be at the Grand Unified scale, 
$\gamma_{ab}$ the string world-sheet metric, and $h$ a 
scalar field.

A linear interaction can be considered quite generally. 
It can also arise if there is a bosonic condensate on the 
string. To see this explicitly, consider the model 
\begin{equation}
S = S_0 [\Phi, H, \ldots ] + 
     \kappa \int d^4 x (\Phi^\dag \Phi - M^2) H^\dag H
\label{ftaction}
\end{equation}
where $S_0$ is a field theory action that yields cosmic
string solutions when $\Phi$ gets a vacuum expectation
value (VEV). $H$ is a scalar field that we can take to be
the electroweak Higgs for concreteness. 

At energy scales above the Grand Unified scale, the VEVs 
of $\Phi$ and $H$ are both zero. At lower energy scales, 
but still above the electroweak scale, $\Phi$ gets a VEV 
so that $| \langle \Phi \rangle |^2 = M^2$ but 
$|\langle H \rangle |^2 =0$. At this stage, we also 
have string solutions whose tension is $\mu \sim M^2$ 
and width is $\sim M^{-1}$. Inside the core of the string, 
where $\Phi^\dag \Phi$ can become small, it may be
favorable for $H$ not to vanish since the coefficient of
the $H^\dag H$ term in Eq.~(\ref{ftaction}) becomes negative. 
As in the case of bosonic superconducting strings 
\cite{Witten:1984eb}, there can be an $H$ condensate in the 
core of the string.
Since $M$ is the only mass scale in the problem at
this stage, the magnitude of $H$ is of order $M$ within 
the core of the string. At a lower energy scale, $H$ 
too gets a VEV. For example, if $H$ is the electroweak
Higgs, this scale is $m \sim 100$ GeV. We will assume
$m \ll M$ and hence the VEV of $H$ within the string
is unaffected by the lower scale (electroweak) symmetry
breaking.
 
The interaction term in Eq.~(\ref{ftaction}) can now
be written as
\begin{eqnarray}
S_{\rm int} &=& 
     \kappa \int d^2 \sigma \int d^2 x_\perp \sqrt{-\gamma}
            (\Phi^\dag \Phi - M^2) H^\dag H \nonumber \\
      &=&
     \kappa \int d^2\sigma \sqrt{-\gamma} \int d^2 x_\perp
            (\Phi^\dag \Phi - M^2) \times \nonumber \\
      && \hskip 1 in (\langle H \rangle_{\rm in} +h )^\dag 
                  (\langle H \rangle_{\rm in} +h )
               \nonumber \\
      &\approx &
    - \kappa M \int d^2\sigma \sqrt{-\gamma} ~ h + \ldots
\end{eqnarray}
In the first line we have split the integral into
world-sheet and transverse integrals and included the
Jacobian factor ($\sqrt{-\gamma}$) where $\gamma_{ab}$ 
($a,b=0,1$) denotes the induced metric on the string 
world-sheet. The subscript ``in'' on the angular brackets 
denotes that the relevant value is the VEV within the core 
of the string and we take $\langle H \rangle_{\rm in} \sim M$. 
Note that $h$ in the last line denotes the radial component 
of $H$ evaluated on the world-sheet.
                          
In the next section, we will estimate the radiation of
Higgses from cusps on cosmic string loops.

\section{Higgs emission}
\label{higgsemission}

The equation of motion for the Higgs field is
\begin{equation}
(\square + m^2) h = j
\end{equation}
where
\begin{equation}
j(x) = - \kappa M \int d\tau ~ d\sigma ~ \sqrt{-\gamma} ~ 
                  \delta^{4} (x-X(\sigma,\tau))
\end{equation}
Then the number of Higgs particles with momentum $k$ 
produced due to a source is 
\begin{equation}
dN_k = | {\tilde j}(\omega_k, {\bf k}) |^2
          \frac{d^3k}{2\omega_k}
\label{dNk}
\end{equation}
where the Fourier transform of the source $j(x)$ is
given by
\begin{equation}
{\tilde j}(\omega_k, {\bf k})
= -\kappa M \int d\tau d\sigma \sqrt{-\gamma} 
       e^{- i [\omega_k \tau - {\bf k} \cdot {\bf X}(\sigma , \tau)]}
\label{tildej}
\end{equation}
where $X^\mu (\sigma , t)$ denotes the string world sheet 
and $\omega_k = \sqrt{k^2 + m^2}$.

The dominant contribution to ${\tilde j}$ comes from
the region around the cusp where 
$| \omega_k \tau - {\bf k} \cdot {\bf X}(\sigma , \tau)| < 1$. 
Choosing world-sheet coordinates so that the cusp occurs at 
$\sigma =0$ and $\tau =0$, this occurs for a range 
$|\tau | ,|\sigma | < L/(kL)^{1/3}$ provided $k > m \sqrt{mL}$
(see Appendix~\ref{loopdynamics}). Also, 
\begin{equation}
\sqrt{-\gamma} = 1-{\dot {\bf x}}^2
\sim \frac{|\sigma |^2}{L^2} \sim  (kL)^{-2/3}
\label{sqrtgamma}
\end{equation}
Therefore
\begin{equation}
{\tilde j}(\omega_k, {\bf k}) \sim \kappa M \frac{|\sigma |^4}{L^2}
 \sim \frac{\kappa ML^2}{(kL)^{4/3}}
\end{equation}
The angular width of the beam of particles emitted from 
the cusp can also be estimated by evaluating the integral 
in Eq.~(\ref{tildej}) in the stationary phase approximation,
or as in Ref.~\cite{Vilenkin:1986zz}. The result is
\begin{equation}
\theta \sim \frac{1}{(kL)^{1/3}}
\label{thetaestimate}
\end{equation}

Then Eq.~(\ref{dNk}) gives,
\begin{equation}
dN_k \sim |{\tilde j}|^2 \theta ^2 k dk 
\sim \kappa^2 M^2 L^{2/3} \frac{dk}{k^{7/3}}
\label{higgs1cusp}
\end{equation}
The estimate applies for 
\begin{equation}
k \in (m\sqrt{mL} , M\sqrt{ML}) \ .
\label{krange}
\end{equation}
The upper cut-off 
on $k$ arises because the wavelength of the emitted
particles should be larger than the string width in the
rest frame of the cusp: $\lambda >  M^{-1}$. Boosting
to the rest frame of the loop, this yields 
$k < M/\sqrt{1-{\dot {\bf x}}^2}$ and together with
the estimate in Eq.~(\ref{sqrtgamma}) gives the upper 
cut-off.  The lower cut-off comes from the 
requirement that $|k\cdot X| < 1$ (see 
Appendix~\ref{loopdynamics}).
For $k < m\sqrt{mL}$, $|\sigma |\sim k/m^2$ and the 
spectrum is a rapidly increasing function of $k$ 
\begin{equation}
dN_k \sim \kappa^2 \frac{M^2 k^{25/3}}{m^{16} L^{14/3}} dk
\end{equation}
Hence $dN_k$ goes to zero very fast as $k \to 0$.
In the following sections, we will ignore this part
of the spectrum and only consider $k > m\sqrt{mL}$.

\begin{figure}
  \includegraphics[width=2.5in,angle=0]{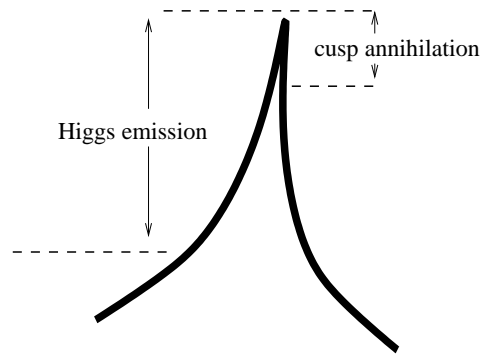}
\caption{Higgs emission with momentum $k$ occurs from a 
section of the string with length $\sim L/(kL)^{1/3}$. 
Cusp annihilation occurs over a length 
$\sim \sqrt{L/M}$ \cite{Olum:1998ag}
and is smaller than the region for Higgs emission for
$k < M \sqrt{ML}$.
}
\label{cusppicture}
\end{figure}

At this stage we can also compare Higgs emission from
cusps to the process of cusp annihilation (see 
Fig.~\ref{cusppicture}). Higgs emission at momentum $k$ 
occurs over a length $L/(kL)^{1/3}$. With the upper bound, 
$k=k_{\rm max} = M\sqrt{ML}$, this length is $\sqrt{L/M}$
and coincides with the cusp annihilation length
\cite{Olum:1998ag}. Hence, cusp annihilation does not 
affect our estimates of Higgs emission for $k < k_{\rm max}$. 

A caveat to this statement is that the presence of the condensate 
should not significantly change the string dynamics. Also,
we have been considering cusp annihilation, but the condensate
itself has some width which is larger than the string width and 
there could, in principle, be ``condensate annihilation'' even 
where there is no cusp annihilation \cite{Vilenkincomment}.

\section{Higgs injection function}
\label{higgsinjection}

Let $d\Phi_H (k,t)$ be the number of Higgs produced by
string cusps with energy $(k,k+dk)$ at time $t$ per
unit volume per unit time. $d\Phi_H$ is called the
``Higgs injection function''. Then
\begin{equation}
d\Phi_H (k,t) = dk \int_{L_{\rm min}}^{L_{\rm max}} dL ~
   \frac{dN_H}{dk}  \frac{dn_L}{dL} \frac{dN_c}{dt}
\label{dPhiH}
\end{equation}
where the first factor in the integrand is the number of
Higgses with energy $k$ produced by a cusp on a loop of
length $L$. The second factor is the number density of
loops of length $L$ at time $t$. The third factor is the
number of cusps per unit time on a loop of length $L$.
The integration is over all loops of length from $L_{\rm min}$
to $L_{\rm max}$. The smallest loop can have length
$\sim M^{-1}$ but most of the ultra-high energy cosmic ray 
signal will come from longer loops. $L_{\rm max}$ is
clearly bounded by the cosmic horizon size $\sim t$ but,
for a fixed value of $k$, it is also bounded due to the 
constraints in Eq.~(\ref{krange}). 

Let us deal with the last factor in (\ref{dPhiH}) first.
Since the motion of the loop is periodic or quasi-periodic
\begin{equation}
\frac{dN_c}{dt} = \frac{f_c}{L}
\end{equation}
where $f_c$ is a parameter that gives the average number of
cusps on a loop per oscillation period. For loops that aren't
too complicated, we expect $f_c \sim 1$.

The rate of Higgs production from a single cusp -- the first 
factor in the integral in (\ref{dPhiH}) -- is given in 
Eq.~(\ref{higgs1cusp}). Now we need to determine the second 
factor -- the number density of loops -- in the integrand
in Eq.~(\ref{dPhiH}). 

The number distribution of cosmic string loops is 
currently under discussion 
\cite{Martins:2005es,Vanchurin:2005pa,Ringeval:2005kr,
Dubath:2007mf,Vincent:1997cx,Rocha:2007ni}
In \cite{Vincent:1997cx} the authors find that loops 
will only be of microscopic size, in which case loops 
decay very quickly. The conventional scenario, though,
is where there is a distribution of loops of all sizes
at all times. Simulations of (non-radiating) string 
networks in an expanding universe give
\begin{equation}
\frac{dn_L}{dL} = \frac{A}{L_i^2 t^2}
\end{equation}
where $A \approx 10$ and $L_i$ is the initial length of
the loop. In these simulations, the loops do not shrink due
to radiation and $L=L_i$ at all times. To include the effects
of radiation from string loops we will express the initial 
length, $L_i$, in terms of the length at time $t$. This is 
given by the differential equation for the rate of energy loss
\begin{equation}
\mu \frac{dL}{dt} = - \Gamma_g G\mu^2 - \Gamma_h \frac{\mu}{\sqrt{mL}}
\label{dLdt}
\end{equation}
where $\Gamma_g$ and $\Gamma_h$ are numerical coefficients
characterizing the gravitational and Higgs radiation. 
The energy lost to Higgs radiation is found by integrating
Eq.~(\ref{higgs1cusp}) after multiplication by $k$. In 
Eq.~(\ref{dLdt}), the energy lost in one cusp event is 
averaged over one oscillation period of the loop to get a 
rate of energy loss.

Eq.~(\ref{dLdt}) has to be solved with the initial condition
$L(t=0) = L_i$, which is equivalent to assuming that all the
string loops are effectively laid down at $t=0$. Strictly,
there will be loops of size $L_i$ that are produced at some
later cosmic time $t_i$, and these will then evolve. However, 
for the purpose of finding the distribution function for
loops, one can view these as larger loops that were produced 
at $t=0$ which then shrunk to $L_i$ at the time $t_i$. This
is the usual scheme for determining the loop distribution
function with the inclusion of loop evaporation (e.g.
see \cite{VilenkinShellard}). A more sophisticated analysis 
that attempts to take radiation backreaction into account 
\cite{Rocha:2007ni} gives a similar loop distribution function 
though with a different dependence on $L$. However, the effects
of backreaction on the loop {\it production} function, which feeds
into the loop {\it distribution} function, has not yet been
included \cite{Rocha:2007ni} and hence we have stayed with 
the conventional loop distribution function.

With the rescalings
\begin{equation}
y = \left ( \frac{\Gamma_g G\mu}{\Gamma_h} \right )^2 m L \ , \ \ 
x = \left ( \frac{\Gamma_g G\mu}{\Gamma_h} \right )^3 \Gamma_h m t
\label{rescalings}
\end{equation}
Eq.~(\ref{dLdt}) becomes
\begin{equation}
\frac{dy}{dx} = -1 - \frac{1}{\sqrt{y}}\ , \ \ 
y(x=0) = y_i
\end{equation}
The differential equation can be solved but it is simpler
to approximate it as
\begin{eqnarray}
\frac{dy}{dx} &=& -1 \ , \ \  y > 1 \nonumber \\
\frac{dy}{dx} &=& -\frac{1}{\sqrt{y}} \ , \ \ y < 1
\label{dybydx}
\end{eqnarray}

\begin{figure}
  \includegraphics[width=3.0in,angle=0]{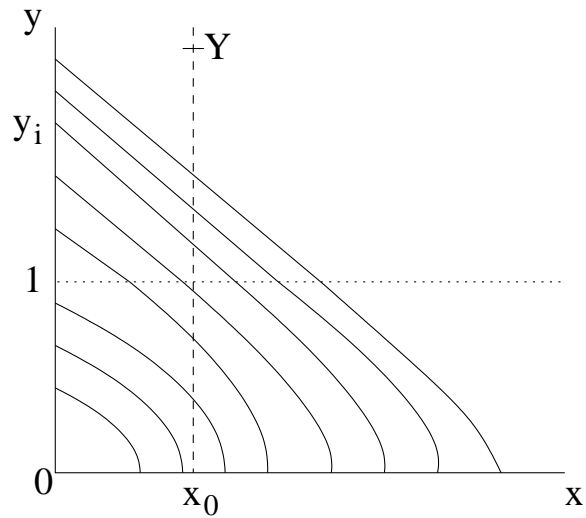}
\caption{The integration in $I(Y,x)$ is over $y$ (dashed vertical
line) at a definite value of $x$ (say $x_0$). The different loop
trajectories are labeled by the initial length, $y_i$, at 
$x=0$. The dotted line at $y=1$ is where the differential
equation changes behavior (see Eq.~(\ref{dybydx})). For small 
values of $y(x_0)$, $y_i \approx x$ up to factors of order 1. 
For large $y(x_0)$, $y_i \approx y$.
}
\label{integrationregion}
\end{figure}

The Higgs injection function in Eq.~(\ref{dPhiH}) can now be 
written as
\begin{equation}
d\Phi_H (k,t) = \frac{\mu A f_c}{t^2} \alpha^{4/3} 
                \frac{dk}{k^{7/3}} I(Y,x)
\label{Hinj}
\end{equation}
where $Y \equiv \alpha L_{\rm max}$ with
\begin{equation}
\alpha \equiv \left ( \frac{\Gamma_g G\mu}{\Gamma_h} \right )^2 m
\label{alpha}
\end{equation}
and
\begin{equation}
I(Y,x) = \int_0^Y \frac{dy}{y^{1/3}} \frac{1}{y_i^2}
\label{Iyx}
\end{equation}
Note that we have set the minimum length to 0 since the 
integration will be dominated by the upper limit.
The maximum length (needed to determine $Y$) cannot be more 
than the Hubble size but it is also restricted by the value 
of $k$ since $k > k_{\rm min} = m \sqrt{mL}$. Therefore
\begin{equation}
Y = \alpha ~ {\rm min} \left ( \frac{k^2}{m^3}, t \right )
\label{Ydefn}
\end{equation}
               
To determine $I(Y,x)$ we need to write $y_i$ in terms of
$y$ and $x$ in Eq.~(\ref{Iyx}). For this we need to solve
Eq.~(\ref{dybydx}) for all $y_i$ to obtain $y(x;y_i)$. 
Then this function should be inverted to get $y_i (y(x),x)$ 
which can then be inserted in Eq.~(\ref{Iyx}) 
(see Fig.~\ref{integrationregion}). We shall do these steps
approximately but analytically.

If $Y$ is very small compared to $x$, the integration over 
$y$ is for $y \ll x$ and then $y_i \approx x$ up to some 
numerical factor of order 1. Therefore
\begin{equation}
I(Y,x) \approx \frac{Y^{2/3}}{x^2} \ , \ \ Y \ll x
\end{equation}
In the opposite limit, $Y \gg x$, the integral
needs to be split into two pieces, one for $y$ from
0 to $x$ and the other from $x$ to $Y$. The first piece
is approximated as for the $Y \ll x$ case. The second piece 
is different since here $y_i \sim y$ is more appropriate.
More explicitly, for $Y \gg x$,
\begin{eqnarray} 
I(Y,x) &=& \int_0^Y \frac{dy}{y^{1/3}} \frac{1}{y_i^2}
           \nonumber \\
       &\approx& \frac{x^{2/3}}{x^2} +
                     \int_x^Y \frac{dy}{y^{7/3}}
           \nonumber \\
       &\approx& \frac{1}{x^{4/3}} 
\end{eqnarray}
Note that $I(Y,x)$ does not depend on $Y$ and, since $Y$ 
is $k$ dependent (Eq.~(\ref{Ydefn})), neither
does $I(Y,x)$ depend on $k$ in this limit.

Therefore
\begin{equation}
d\Phi_H (k,t) \approx \frac{\mu A f_c}{t^2} \alpha^{4/3} 
                \frac{dk}{k^{7/3}} 
                     \frac{Y^{2/3}}{x^2} \ , \ \ Y \ll x
\end{equation}
and
\begin{equation}
d\Phi_H (k,t) \approx \frac{\mu A f_c}{t^2} \alpha^{4/3} 
                \frac{dk}{k^{7/3}} 
                     \frac{1}{x^{4/3}} \ , \ \ Y \gg x
\end{equation}
A more rigorous derivation will lead to a smooth interpolation
between these asymptotic forms. For the purpose of our
estimates, it is sufficient to extend the above asymptotic
forms so that they connect continuously at $Y=x$.


These equations can be written more neatly in terms of
\begin{equation}
k_* \equiv m \sqrt{\Gamma_g G\mu m t}
\label{kstar}
\end{equation}
Then
\begin{eqnarray}
d\Phi_H (k,t) &=&\frac{\mu A f_c}{t^2}  
      \frac{m^4}{k_*^4} \frac{dk}{k} \ ,  \ \ k \le k_* \\
d\Phi_H (k,t) &=&\frac{\mu A f_c}{t^2}  
      \frac{m^4}{k_*^{8/3}} \frac{dk}{k^{7/3}} \ ,
                                         \ \  K > k \ge k_* 
\label{dPhilargek}
\end{eqnarray}
where $K \equiv m \sqrt{m t}$ and the bound $k < K$ ensures 
that $Y =\alpha k^2/m^3$ {\it i.e.} the loops are less than
the horizon size. However, the $k^{-7/3}$ fall off is
also valid for $k > K$ because then $I(Y,x)$ in Eq.~(\ref{Hinj})
is independent of $k$. So the restriction $K > k$ in 
Eq.~(\ref{dPhilargek}) may be dropped. Eq.~(\ref{dPhilargek})
is our estimate for the Higgs injection function.

An important feature of the injection function in
Eq.~(\ref{dPhilargek}) is that it is inversely proportional
to the string scale. Thus lighter strings inject more Higgses
than heavier strings. This feature was also noted in 
Ref.~\cite{Bhattacharjee:1989vu} and can be explained by 
the greater longevity of light string loops. 

\section{Diffuse photon flux}
\label{cascade}

Once Higgses are injected into the cosmological medium, 
they will decay, lose energy and eventually cascade into 
gammas in the EGRET energy range. The energy in gammas
in the EGRET window is estimated as a fraction of the
total injected energy using \cite{Berezinskybook}
\begin{equation}
\omega_{\rm cas} \equiv \frac{f_\pi}{2} \int_0^{t_0} dt
                  \int dk ~ k \frac{d\Phi_H}{dk} 
                  \frac{1}{(1+z)^{4}}
\end{equation}
where $f_\pi$ is the fraction of energy of the Higgses
that goes into pions and $z$ is the cosmological redshift. 
The 1/2 accounts for how much energy goes into gammas.
The diffuse gamma ray background measured by EGRET is 
on the order of
$\omega_{\rm cas,obs} \sim 10^{-6}\ {\rm eV/{{cm}^3}}$ 
in the energy range 10 MeV-100 GeV \cite{Sreekumar:1997un}.

With the Higgs injection function found in the previous
section, we have
\begin{eqnarray}
\int dk k \frac{d\Phi_H}{dk} &\approx &
 \int_0^{k_*} dk \frac{\mu A f_c}{t^2} \frac{m^4}{k_*^4}
                 \nonumber \\
&\sim& \frac{\mu A f_c m}{(\Gamma_g G\mu m)^{3/2}} 
                 \frac{1}{t^{7/2}}
\end{eqnarray}
The integration is dominated by the upper limit ($k_*$).
The integration for $k \in (k_*,\infty)$ will be dominated
by its lower limit because of the faster ($k^{-4/3}$) fall
off and will give a comparable result.

Now with $1+z = a(t_0)/a(t)$ where $a(t)$ is the scale factor in
a matter dominated universe the time integration can be done
and gives
\begin{equation}
\omega_{\rm cas} \sim \frac{3Af_\pi f_c}{(\Gamma_g G\mu )^{3/2}}
          \frac{\mu}{m^2} (m t_0)^{1/2} \frac{m}{t_0^3}
\end{equation}
The time integral is dominated by the present cosmic time and
Higgs production during the radiation dominated era has been
ignored. 
As noted at the end of the previous section, the energy 
cascade into gamma rays is larger for lighter strings.

To get a feel for the numbers, we use $m=10^2 m_2$ GeV,
$\mu = (10^{13} \eta_{13}{\rm GeV})^2$ (therefore 
$G\mu = 10^{-12} \eta_{13}^2$), $t_0 =10^{28}$ cm. 
Also we take $A\sim 10$, $\Gamma_g \sim 100$
and we assume $f_\pi \sim 1$. Then 
\begin{equation}
\omega_{\rm cas} \sim 10^{-13} \eta_{13}^{-1} m_2^{-1/2}
                       \frac{\rm eV}{\rm cm^3}
\end{equation}
This is quite a bit smaller than the EGRET observation
unless the string scale is less than $\sim 10^6$ GeV.
Lighter strings with the interactions in Eq.~(\ref{ftaction}) 
are ruled out, though with the caveat that the dynamics of
much lighter strings ($M < 1$ TeV) can be dominated by 
friction until the present cosmic epoch and so the network 
properties and loop dynamics can be quite different 
\cite{VilenkinShellard}.

\section{Diffuse proton flux}
\label{protonflux}

We shall follow Berezinsky et al \cite{Berezinsky:1998ft} 
and use a power-law fragmentation function to obtain the 
diffuse proton flux, $I_p (E)$, as (see Eq.~(A13) of
\cite{Berezinsky:1998ft})
\begin{equation}
I_p(E) = \frac{(2-p)f_N}{4\pi p} \frac{{\dot n}_H}{E_H}
         \left ( \frac{E}{E_H} \right )^{-p} R_p (E)
\end{equation}
Here $p=1.9$ \cite{Aloisio:2003xj}
and $f_N$ is the fraction of energy transferred 
to nucleons when a Higgs decays which, for order of magnitude 
estimates, we will take to be 1. ${\dot n}_H$ is the rate at 
which Higgses are being produced per unit volume and can be
found from our result for $d\Phi_H$ above. $E_H$ is the
energy at which the Higgses are produced and $R_p (E)$
is the proton attenuation length at energy $E$ due to
scattering off the CMB. At energies of $10^{19}$ eV 
{\it i.e.} below the GZK cut-off, 
$R_p \sim t_0 \sim 10^{28}$ cm.

To estimate $I_p (E)$ we take
\begin{equation}
E_H = k_* \approx 10^{19} \eta_{13} m_2^{3/2} ~ {\rm GeV}
\end{equation}
and
\begin{equation}
{\dot n}_H (k_*,t_0) = \frac{\mu A f_c}{t_0^2} \frac{m^4}{k_*^4}
 \approx 10^{-58} m_2^{-2} \eta_{13}^{-2} ~ 
               {\rm cm^{-3} ~ s^{-1} sr^{-1}}
\end{equation}
This gives
\begin{equation}
E^3 I_p (E) \approx 10^{18} 
       \left ( \frac{E}{E_{19}} \right )^{1.1}
       {\rm m_2^{-0.65} \eta_{13}^{-1.1}}  
{\rm \frac{eV^2}{m^2-s-sr}}
\end{equation}
where $E_{19} = 10^{19} {\rm eV}$. This is to be compared
with the observed flux \cite{Abraham:2008ru}
\begin{equation}
[E_{19}^3 I_p (E_{19}) ]_{\rm obs} \approx 10^{24} 
{\rm \frac{eV^2}{m^2-s-sr}}
\end{equation}
Hence the proton flux is much less than the observed flux
unless $\eta_{13} \sim 10^{-6}$, in which case both the
diffuse gamma flux and the diffuse proton flux are comparable 
to observations.

\section{Direct flux}
\label{directflux}

\begin{figure}
\includegraphics[width=2.5in,angle=0]{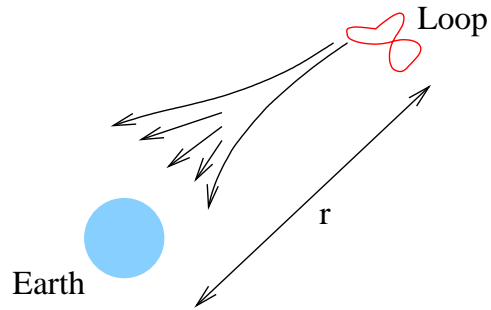}
\caption{A cusp on a loop close to Earth emits particles 
that lose energy as they propagate and also spread out.
}
\label{earthloop}
\end{figure}
String loops that are relatively close-by to the
Milky Way may beam Higgses directly at the Earth and
these would be seen as ultra-high energy cosmic rays.
However, as the particles propagate from the cusp to
the Earth, they lose energy due to scattering with 
various components of the cosmological medium. The
energy of a particle drops exponentially with distance
from the cusp. If $k'$ is the energy of a particle 
at the cusp, $k$ is the energy at Earth, and $r$ is 
the distance from the loop to the Earth, we have
\begin{equation}
k = k' e^{-r/R}
\label{kk'}
\end{equation}
As the initially emitted Higgs particle decays and
loses energy due to interactions, the products 
spread out over a wider beam angle, $\theta_b$.
The total energy in all the particles is proportional
to $k \theta_b^2$ and we assume that this remains
roughly constant along the length of the beam.
Therefore the angular spread of the beam at Earth
is 
\begin{equation}
\theta_b^E = \theta_b^l e^{+r/2R}
\label{thetaEthetal}
\end{equation}
where the superscripts denote Earth ($E$) and the loop
($l$).  The number of particles at Earth with energy $k$
produced from a loop of length $L$, at distance $r$ 
follows from Eq.~(\ref{Hinj}) 
\begin{equation}
d\Phi^E (k,L,r) = \frac{\mu A f_c \alpha^{4/3}}{t^2}
                   \frac{dk'}{k'^{7/3}}
                    \frac{dy}{y^{1/3}} \frac{1}{y_i^2}
                    dV ~ d\Omega_b
\end{equation}
Note that the left-hand side is the flux at Earth and 
hence is at energy $k$, while the right-hand side contains
the injected flux at the location of the loop and hence
is at $k'$.

The beaming solid angle is
\begin{equation}
d\Omega_b \sim \pi ({\theta_b^E})^2 
            = \pi ({\theta_b^l})^2 e^{+r/R}
            = \frac{\pi e^{r/R}}{(k' L)^{2/3}}
\end{equation}
where we have used Eq.~(\ref{thetaestimate}).
Using Eq.~(\ref{kk'}) and the rescalings in 
(\ref{rescalings}) and integrating over loop lengths
and spatial volume, we get
\begin{equation}
d\Phi^E (k) = \frac{4\pi^2 \mu A f_c \alpha^2}{t^2}
   \frac{dk}{k^3} \int_0^\infty dr r^2 e^{-r/R}
   \int_{y_{\rm min}}^{y_{\rm max}} \frac{dy}{y} \frac{1}{y_i^2}
\label{dPhi}
\end{equation}

The limits of integration over loop lengths ($y$) also
depend on $r$ since (see Eq.~(\ref{krange}))
\begin{equation}
y_{\rm min} = \alpha \frac{{k'}^2}{M^3} 
            = \alpha \frac{k^2}{M^3} e^{2r/R}
\end{equation}
\begin{equation}
y_{\rm max} 
 = \alpha ~ {\rm min}\left ( \frac{{k'}^2}{m^3} , t \right )
 = \alpha ~ {\rm min} \left ( \frac{k^2}{m^3} e^{2r/R}, t \right )
\end{equation}
where $\alpha$ is defined in Eq.~(\ref{alpha}). 

\begin{figure}
\includegraphics[width=3.0in,angle=0]{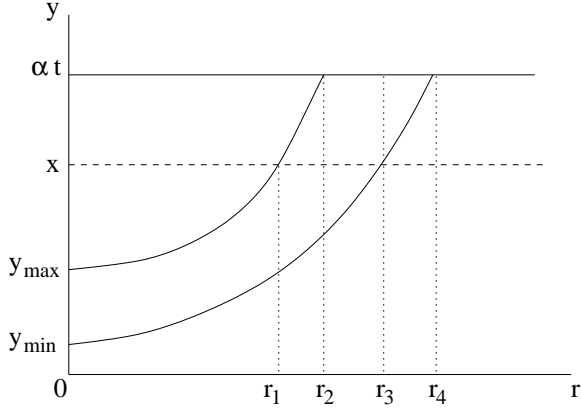}
\caption{The integration region in (\ref{dPhi}) in 
the $ry-$plane falls in the region bounded by the $y_{\rm min}$
and $y_{\rm max}$ curves from $r=0$ to $r=r_4$ while
the integrand depends on whether $y < x$ or $y > x$.
The dominant contribution to the integral comes
from the region $0< r < r_1$.
}
\label{ryregion}
\end{figure}

In the $(r,y)$ plane of integration in Eq.~(\ref{dPhi}),
there are four values of $r$ which are significant.
The first, $r_1$, is where $y_{\rm max}$ becomes equal
to $x$ (see Eq.~(\ref{rescalings})), and is given by
\begin{equation}
e^{-r_1/R} = \frac{k}{k_*}
\end{equation}
The second, $r_2$, is where $y_{\rm max}$ becomes equal to
the Hubble size loop
\begin{equation}
e^{-r_2/R} = \sqrt{\Gamma_h G\mu} \frac{k}{k_*}
\end{equation}
The third, $r_3$, is where $y_{\rm min}$ becomes equal
to $x$
\begin{equation}
e^{-r_3/R} = \left ( \frac{m}{M} \right )^{3/2} \frac{k}{k_*} 
\end{equation}
and the last, $r_4$, where $y_{\rm min}$ equals $\alpha t$
\begin{equation}
e^{-r_4/R} = \sqrt{\Gamma_h G\mu} \left ( \frac{m}{M} \right )^{3/2}
                               \frac{k}{k_*}
\end{equation}

The integration in the $ry$-plane in Eq.~(\ref{dPhi}) can
be split up into four integrations over the ranges $(0,r_1)$,
$(r_1,r_2)$, $(r_2,r_3)$ and $(r_3,r_4)$. For $y$ values less
than $x$ we can use $y_i \sim x$, as argued in 
Sec.~\ref{higgsinjection}.
For $y$ values greater than $x$, we use $y_i \sim y$.
While the full integration can be done, it is unnecessary
because the dominant contribution comes from the interval
$(0,r_1)$ {\it i.e.} the closest strings. The other integrals 
are suppressed by powers
of $k/k_*$ (see Eq.~(\ref{kstar})). The integration over
$(0,r_1)$ is 
\begin{eqnarray}
I_1 &=& \int_0^{r_1} dr ~ r^2 e^{-r/R} 
        \int_{y_{\rm min}}^{y_{\rm max}} \frac{dy}{y} \frac{1}{y_i^2}
        \nonumber \\
    &=& \frac{6}{x^2} \ln \left ( \frac{M}{m} \right ) R^3
\end{eqnarray}
where we have used $y_i \approx x$, 
$y_{\rm max} /y_{\rm min} = (M/m)^3$, and assumed $k/k_* \ll 1$.

This leads to 
\begin{equation}
k \frac{d\Phi^E}{dk} = 24\pi^2 A f_c ~ \ln \left( \frac{M}{m} \right )
          \frac{M^2}{(\Gamma_h G\mu )^2 k^2} \frac{R^3}{t^4}
\label{directfluxeq}
\end{equation}

Using $t=10^4 ~ {\rm Mpc} = 10^{17}$ s, and dividing by the Earth's
surface area, $10^9 ~ {\rm km^2}$, we get the flux per unit area
\begin{eqnarray}
k \frac{d\Phi^E}{dA dk} &\approx& 
10^{-3} \left ( \frac{10^{20} ~ {\rm eV}}{k} \right )^2 
\left ( \frac{10^7 ~ {\rm GeV}}{M} \right )^2 
\nonumber \\
&& 
      \times \left ( \frac{R}{5 ~{\rm Mpc}} \right )^3 
~ {\rm m^{-2} ~ s^{-1}}
\end{eqnarray}
For $M \sim 10^{13}$ GeV, this is comparable to the observed 
flux of ultra high energy cosmic rays,
$10^{-36} {\rm m^{-2}s^{-1}sr^{-1}eV^{-1}}$ at 
$10^{20} {\rm eV}$ \cite{Abraham:2008ru}. Note that 
$m$ only enters through the logarithm and its precise value 
does not make much difference to the overall estimate.

The direct photon flux will also be given by Eq.~(\ref{directfluxeq}),
but the attenuation length $R$ will be specific to photons
which is less than that for protons at energy $10^{19}$ eV,
and the branching ratio for Higgs decay into protons differs
from that to photons. Hence the photon to proton ratio is
\begin{equation}
\frac{\gamma}{p} = \frac{N_\gamma}{N_p} 
          \left ( \frac{R_\gamma}{R_p} \right )^3
\end{equation}
where $N_\gamma$ and $N_p$ are the number of gammas and the
number of protons produced by a Higgs. From Figs. 9 and 11
of Ref.~\cite{Bhattacharjee:1998qc}, and also see
Fig.~2 of \cite{Berezinsky:1998ft},
we find $R_\gamma/R_p \sim 10^{-2}$ at $10^{19}$ eV. 
If the heavy particle emitted by the cusp is the electroweak 
Higgs, the decay products contain a pion fraction of $\sim 0.75$
and a nucleon fraction $\sim 0.15$. The pions then decay into
gamma photons. As a conservative estimate of the number of photons 
to protons in the decay products, we take $N_\gamma/N_p \sim 10^2$. 
Then the gamma to proton fraction at $10^{19}$ eV is $\sim 10^{-4}$, 
which is consistent with the observed AUGER bound 
$\gamma/p < 0.02$ \cite{Collaboration:2009qb}.  As pointed out in 
Ref.~\cite{Berezinsky:1998ft}, at higher energies, since $R_p$ 
falls quite rapidly, we expect the gamma to proton ratio to be 
larger. For example, at $10^{20}$ eV, $R_\gamma/R_p \sim 10^{-1}$ 
and $\gamma /p \sim 10^{-1}$, whereas the observed AGASA and
Yakutsk bound is $0.3$ \cite{Rubtsov:2006tt}. 

We would like to point out that we have performed our analysis 
by taking $R$ to be a constant, as in Eq.~(\ref{kk'}). This
is only an approximation since $R$ depends on the energy of 
the particle and changes as the particle propagates. A more 
complete analysis would take the energy loss to be given by
\begin{equation}
\frac{dk}{dr} = - \frac{k}{R(k)}
\end{equation}
where $R(k)$ follows by considering the various interactions
that a cosmic ray may encounter en route. Our results are
valid only if 
\begin{equation}
\frac{dR}{dk} \biggr | _{k_*}  \ll \frac{R(k_*)}{k_*}
\end{equation}
An improved analysis should take the detailed form of
$R(k)$ into account.

\section{Conclusions}
\label{conclusions}

It is generally difficult to construct astrophysical
scenarios that can accelerate protons to high energies
sufficient to arrive as ultra-high energy cosmic
rays. Top down models involving topological defects can 
naturally give the requisite energies. Depending on the
precise variety of defect, however, they are likely to suffer
from an excess of diffuse gamma ray production.

We have re-visited cosmic ray production from cosmic strings, 
taking into account the possibility of a linear interaction
between a scalar field and the string world-sheet. Such an
interaction arises, for example, if a scalar field acquires
a VEV (``condenses'') within the string. Then beams of Higgs 
particles are emitted from cusps on cosmic string loops. Our 
analysis shows that such events can be responsible for the 
production of UHECR within reasonable parameters and also be 
consistent with measurements of the diffuse backgrounds and 
the photon to proton fraction in UHECR. 

Our results are also interesting because they show that 
strings with bosonic condensates are allowed in a narrow 
window of energy scales. If they are much lighter than 
$\sim 10^{13}$ GeV, they will produce an {\it excess} of UHECR 
and diffuse fluxes, while if they are much 
heavier, they will cause conflicts with other cosmological 
constraints that depend on their gravitational effects. 
However, if the strings are around the Grand Unified scale,
they can produce the observed flux of ultra high energy
cosmic rays and also be heavy enough to be detected by 
their gravitational signatures in the near future. 

We have focused on deriving {\it analytical} estimates 
of cosmic ray fluxes under various simplifying assumptions. 
Further work is needed to obtain more detailed estimates 
that can be compared to observations.

\begin{acknowledgments}
I am very grateful to Nima Arkani-Hamed, Clifford Cheung, and 
Jared Kaplan for early collaboration and for continued support
and encouragement. I am also grateful to Venyamin Berezinsky, 
Francesc Ferrer, Uri Keshet, Shmuel Nussinov, Ken Olum, Grisha
Rubtsov, Alex Vilenkin, Jay Wacker and Edward Witten for their 
comments and advice.  This work was supported by the U.S. Department 
of Energy at Case Western Reserve University.
\end{acknowledgments}

\appendix

\section{Loop dynamics and cusps}
\label{loopdynamics}

Here we summarize some known features of cosmic string
loops that are used to obtain the radiation estimates in 
Sec.~\ref{higgsemission}.

A cosmic string loop oscillates under its own tension such that
its world-sheet can be written as
\begin{equation}
{\bf X}(\sigma ,t ) = \frac{1}{2} [ {\bf a}(\sigma -t)
                                   + {\bf b}(\sigma +t) ]
\end{equation}
where ${\bf a}$ and ${\bf b}$ satisfy
\begin{equation}
|{\bf a}'| =1 = |{\bf b}'|
\end{equation}
where primes denote derivatives with respect to the argument.
Also, since the loop is a closed string,
\begin{equation}
\int_0^L {\bf a}' d\sigma = 0 = \int_0^L {\bf b}' d\sigma   
\end{equation}
Therefore $-{\bf a}'$ and ${\bf b}'$ are two closed curves on a
unit sphere whose centers of mass coincide with the center of
the sphere. These curves generally intersect leading to 
\begin{equation}
{\bf a}' (\sigma -t) = - {\bf b}' (\sigma +t)
\end{equation}
for one or more values of $\sigma$ and $t$. Since the velocity
of a point on the loop is
\begin{equation}
{\bf v}(\sigma ,t ) = \frac{1}{2} [ - {\bf a}' + {\bf b}' ] \ ,
\end{equation}
intersection of the $-{\bf a}'$ and ${\bf b}'$ curves
implies a point on the loop that reaches
the speed of light at one instant per oscillation. Such a point 
on the string is called a ``cusp''. The ultra-high boost factors
at the cusp can be responsible for burst-like events from cosmic
string loops.

We choose our string parametrization so that the cusp
occurs at $\sigma=0=t$, and then expand the functions 
${\bf a}$ and ${\bf b}$ around a cusp 
\begin{equation}
{\bf a}(\zeta_-) = {\bf a}_0' \zeta_- + 
               \frac{1}{2} {\bf a}_0'' \zeta_-^2
             + \frac{1}{6} {\bf a}_0''' \zeta_-^3 + \ldots
\end{equation}
\begin{equation}
{\bf b}(\zeta_+) = {\bf b}_0' \zeta_+ + 
               \frac{1}{2} {\bf b}_0'' \zeta_+^2
             + \frac{1}{6} {\bf b}_0''' \zeta_+^3 + \ldots
\end{equation}
where $\zeta_\pm = \sigma \pm t$.
The expansion coefficients are constrained by
\begin{eqnarray}
{\bf a}_0' &=& - {\bf b}_0' \ ; \ \ 
{\bf a}_0'\cdot {\bf a}_0'' = {\bf b}_0'\cdot {\bf b}_0'' =0 
\nonumber \\
|{\bf a}_0'| &=& |{\bf b}_0'| = 1 \ .
\end{eqnarray}

These expansions give
\begin{equation}
\omega_k t - {\bf k} \cdot {\bf X}
\sim ( \sqrt{k^2+m^2} -k ) \zeta + \delta ~ k L^{-2} \zeta^3
          + \ldots
\label{phase}
\end{equation}
where we have taken $\zeta_+ \sim \zeta_-$ and denoted them
collectively by $\zeta$.
$\delta$ is a dimensionless constant that depends on
the shape of the cusp. We will take $\delta \sim 1$.
To have
$|\omega_k t - {\bf k} \cdot {\bf X}| < 1$, we require
that both terms in Eq.~(\ref{phase}) be less than 1. 
This gives $\zeta < {\rm min} ( k/m^2, L (kL)^{-1/3} )$. 
For $k > m \sqrt{mL}$, we get $\zeta < L (kL)^{-1/3}$
and for $k <  m \sqrt{mL}$, we get $\zeta < k/m^2$.

\end{document}